\documentclass{ws-p8-50x6-00}

\makeatletter
\newcommand\preprintnote{%
        To appear in Proceedings of 9th International Workshop on Deep
        Inelastic Scattering and QCD (DIS 2001), Bologna, Italy, 27 Apr--1
        May 2001 
}
\newcommand\preprint{
   \def\ps@plain{%
       \let\@mkboth\@gobbletwo
       \let\@oddhead\@empty
       \def\@oddfoot{%
           \lower1pc\hbox{\fbox{%
              \parbox{\hsize}{{\it \preprintnote}}%
           }}
       }%
   }
   \def\ps@simple{%
       \let\@mkboth\@gobbletwo
       \let\@oddhead\@empty
       \def\@oddfoot{\hfill {\bf \thepage}\hfill}%
   }
   \AtBeginDocument{
       \pagestyle{simple}
   }
}
\makeatother
\preprint

\begin{document}

\title{PDF uncertainties: 
   A strong test of goodness of fit to multiple data sets}

\author{John Collins\footnote{On leave from:
        Physics Department,
        Penn State University, %\\
        104 Davey Laboratory,
        University Park PA 16802
        U.S.A.
     }
}

\address{%
   DESY, Notkestra{\ss}e 85, D-22603 Hamburg, Germany, \\
   {\em and}\\
   II Institut f{\"u}r Theoretische Physik, Universit{\"a}t Hamburg, \\{}
   Luruper Chaussee 149, D-22761 Hamburg, Germany
\\ 
E-mail: collins@phys.psu.edu}

\author{Jon Pumplin}

\address{%
        Department of Physics and Astronomy,\\
        Michigan State University,
        East Lansing MI 48824,
        U.S.A.
\\
E-mail: pumplin@pa.msu.edu}  

\maketitle

\abstracts{%  
   We present a new criterion for the goodness of global fits.  It
   involves an exploration of the variation of $\chi^2$ for subsets of
   data.  
}

%==================================================================
\section{Introduction}

This talk addresses the questions of quantifying how good global fits
are, and of how we would know a theory is wrong; it summarizes our
Ref.~\cite{Collins:2001es}. 

The obvious criterion is that of hypothesis testing: $\chi^2 = N \pm
\sqrt{2N}$ for a good fit with $N$ degrees of freedom.  One should also
apply the same criterion to subsets of data (e.g., from a particular
experiment or reaction), for which the normal range is $\chi_i^2 = N_i \pm
\sqrt{2N_i}$.

In fact, a much stronger criterion applies. The idea was discovered by
contrasting the criteria for a one-standard-deviation effect in hypothesis
testing and in parameter fitting.  When fitting a single parameter, the
one-sigma range of the parameter is found by increasing $\chi^2$ one unit
above its minimum.  But the fit is good (hypothesis-testing) at the
one-sigma level if $\chi_{\rm min}^2$ is within $\sqrt{2N}$ of its normal
value $N$.

What is shown in Ref.~\cite{Collins:2001es} is that the goodness of a
global fit is better tested by applying the parameter-fitting criterion in a
certain way to subsets of data.  This can be much more stringent than
the obvious hypothesis-testing criterion.

%==================================================================
\section{Scenario}

Suppose we have two pre-existing good global fits of parton densities,
called CTEQ and MRST, and that new data arrive from two experiments, TEV
and HERA.  Assume that the $\chi^2$s are as in Table \ref{table:scenario},
so that by the hypothesis-testing criterion, each set of pdf's gives a
good fit to each experiment.

\begin{table}
\caption{Hypothetical comparison of data and pdf's.}
\label{table:scenario}
\centering
  \begin{tabular}{|r|ccc|}
     \hline
              &    TEV      &    HERA      &    Total     \\
      PDF     & (100 pts) & (100 pts) & (200 pts) \\
     \hline
      CTEQ    &      85           &       115         &       200         \\
      MRST    &     115           &        85         &       200         \\
     \hline
  \end{tabular}
\end{table}

In fact we may have a bad fit.  This can be seen by constructing pdf's that
interpolate between CTEQ and MRST, $f_p = p f^{\rm CTEQ} + (1-p) f^{\rm
MRST}$, and then fitting the interpolating parameter $p$.  If, for
example, we have $\chi_{\rm TEV}^2 = 85 + 30(1-p)^2$ and $\chi_{\rm
HERA}^2 = 85 + 30p^2$, then the TEV data implies $p=1\pm 0.18$, while the
HERA data implies $p=0\pm 0.18$.

By converting the problem to one of parameter fitting, we have found that
the theory and experiments are mutually inconsistent in this case, by about
$4\,\sigma$.  If the forms for $\chi^2$ are different, it is possible to
have a good fit to both experiments, but only if neither CTEQ nor MRST fit
the data.  The decreases of 30 in $\chi^2$ between the CTEQ and MRST
values are sufficient to show that at least one of these situations arises.

%==================================================================
\section{General procedure and application to CTEQ5}

Observe that in the hypothetical scenario, CTEQ alone obtains a good 
$\chi^2$, and we only saw a problem
when we brought in MRST.  How can CTEQ alone determine that there is a
problem 
without MRST's assistance, and vice versa?  And how can this be done without
knowing which of $\sim30$ parameters is the important one?

\begin{figure}
  \centering
  \includegraphics[scale=0.47]{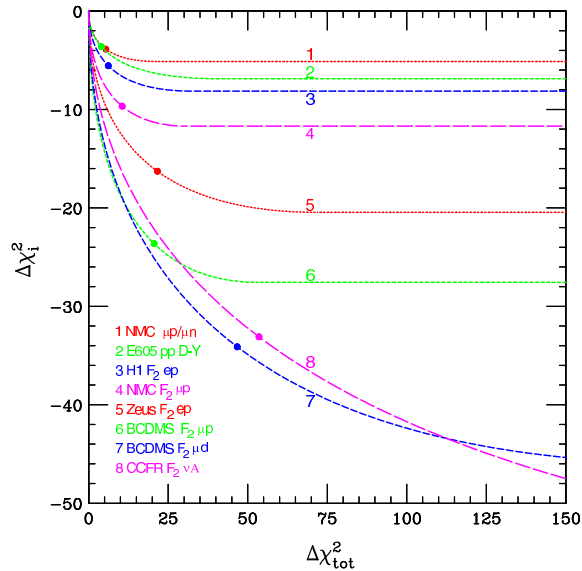}
  \caption{Application of our method to the 8 data sets that have the
     lion's share of the data points used in the
     CTEQ5 \protect\cite{Lai:2000wy} analysis.  The values of
     $\Delta\chi^2$ plotted are the deviations of the $\chi^2$ from the
     values at the overall best fit.  
  } 
  \label{fig:fit30a}
\end{figure}

\begin{table}
\caption{Values of $\chi^2$ for data contributing to the CTEQ5 fit.}
\label{table:CTEQ5}
\centering
\begin{tabular}{|l|rrrr|}
\hline
Expt             &  $N$  & $\chi^2$ & $\chi^2/N$ &
$\frac{\chi^2-N}{\sqrt{2N}}$ \\ 
\hline
1. NMC D/H       &  123  &  111     & 0.90       & -0.8 \\
2. E605          &  119  &   92     & 0.77       & -1.8 \\
3. H1 $F_2$ '96  &  172  &  108     & 0.63       & -3.4 \\
4. NMC H         &  104  &  108     & 1.04       &  0.3 \\
\hline
5. ZEUS $F_2$ '94&  186  &  249     & 1.34       &  3.3 \\
6. BCDMS H       &  168  &  146     & 0.87       & {\bf -1.2}    \\
7. BCDMS D       &  156  &  222     & 1.42       &  3.7 \\
8. CCFR $F_2$    &   87  &   74     & 0.85       & {\bf -1.0} \\
\hline
\end{tabular}
\end{table}

The procedure we propose \cite{Collins:2001es} is as follows:
\begin{itemize}
\item Take pre-determined subsets of data (e.g., experiments).
\item Explore a region of parameter-space with $\chi^2 - \chi^2_{\rm min}$
   up to about $\sqrt{2N}$, e.g., $50$ for CTEQ/MRST, and find the minimum
   of $\chi_i^2$ for each subset.  
\item If $\min(\chi_i^2) - \chi_i^2({\rm global~fit})$ is less than a few 
   units, then experiment $i$ disagrees with global fit.
\item In computing significance, allow for the number of experiments and the
   effective number of parameters determined by each subset of data.  
\end{itemize}
The minimization of $\chi_i^2$ for a given $\chi_{\rm total}^2$ can be
readily implemented by a Lagrange multiplier method:  For each value of a
parameter $\lambda$, minimize
$\chi_{\rm tot}^2({\bf p}) + (\lambda - 1) \chi_i^2({\bf p})$.
The result gives a curve for $\chi_i^2$ as a function of 
$\chi_{\rm tot}^2$.   

The results of applying this procedure to the CTEQ5 fit are shown in Fig.\
\ref{fig:fit30a}. Several of the data sets can be seen to be bad fits,
notably from CCFR and BCDMS.  The criterion is that the $\chi^2$ for the
subset of data decreases by too many units as $\chi_{\rm tot}^2$
increases.  A small decrease is normal and expected.  The bad fit happens
even though nothing exceptional happens according to the
hypothesis-testing criterion, as can be seen in the last column of Table
\ref{table:CTEQ5}.

MRST's plots of $\chi^2$ against $\alpha_s(M_Z)$ in Fig.\ 21 of
Ref.~\cite{Martin:1998sq} independently confirm our physics conclusion,
that current data and QCD theory (including the approximations of NLO 
calculations, neglect of nuclear target and higher-twist corrections, etc.) 
are not in good agreement.

%==================================================================
\section{Conclusions}

We have shown that the quality of a global fit is correctly determined by
testing the variation of $\chi^2({\rm subset})$ for subsets of data as
parameters are varied. A substantial decrease is a symptom of a bad fit, and
the parameter-fitting criterion is the correct one here.  With this method 
small data sets do not get lost compared with the other $\sim1000$
points. The current CTEQ5 global fit appears not to fit the data, and the 
same appears to apply to the MRST fit.

Statistical analysis alone cannot tell us the explanation of this
inconsistency.  Only a physics-based analysis can decide if the problem is
in one of the experiments, if there is a technical error in a theory
calculation, or if there is really new physics that has been measured.
The statistics only give a diagnosis of where further investigation will
be most useful.

%==================================================================
\section*{Acknowledgments}

This work was supported in part by the U.S. Department of Energy and by
the National Science Foundation.  We would like to thank R. Thorne for
conversations at this workshop.  JCC would like to thank the Alexander von
Humboldt foundation for an award.

%==================================================================

\end{document}